# A high dynamic-range instrument for SPICA for coronagraphic observation of exoplanets and monitoring of transiting exoplanets

K. Enya*[a], L. Abe[b], S. Takeuchi[a], T. Kotani[a], T. Yamamuro[c]

[a]Institute of Space and Astronautical Science, Japan Aerospace Exploration Agency,
3-1-1 Yoshinodai, Chuo-ku, Sagamihara, Kanagawa 252-5210, Japan;
[b]Laboratoire Hippolyte Fizeau, UMR 6525 Université de Nice-Sophia Antipolis,
Parc Valrose, F-06108 Nice, France;
[c]Optcraft, 3-26-8 Aihara, Sagamihara, Kanagawa 229-1101, Japan

**ABSTRACT**

This paper, first, presents introductory reviews of the Space Infrared Telescope for Cosmology and Astrophysics (SPICA) mission and the SPICA Coronagraph Instrument (SCI). SPICA will realize a 3m class telescope cooled to 6K in orbit. The launch of SPICA is planned to take place in FY2018. The SPICA mission provides us with a unique opportunity to make high dynamic-range observations because of its large telescope aperture, high stability, and the capability for making infrared observations from deep space. The SCI is a high dynamic-range instrument proposed for SPICA. The primary objectives for the SCI are the direct coronagraphic detection and spectroscopy of Jovian exoplanets in the infrared region, while the monitoring of transiting planets is another important target owing to the non-coronagraphic mode of the SCI. Then, recent technical progress and ideas in conceptual studies are presented, which can potentially enhance the performance of the instrument: the designs of an integral 1-dimensional binary-shaped pupil mask coronagraph with general darkness constraints, a concentric ring mask considering the obscured pupil for surveying a wide field, and a spectral disperser for simultaneous wide wavelength coverage, and the first results of tests of the toughness of MEMS deformable mirrors for the rocket launch are introduced, together with a description of a passive wavefront correction mirror using no actuator.

## 1. INTRODUCTION

The detailed study of extra-solar planets (exoplanets) is considered to be one of the most important tasks in space science in the near future. Since the first report by Mayor and Queloz (1995)[1] using the radial velocity method, more than 550 exoplanets have been discovered.[2] It has also been shown that observations monitoring the transits of exoplanets provide a valuable means for studying them.[3] Not only detection but also spectroscopic studies of transiting exoplanets have been carried out.[4-7] The spatially resolved direct detection of an exoplanet by coronagraphic imaging was finally reported.[8,9,10] Recently, the KEPLER mission has vastly extended the number of transiting exoplanet candidates.[11,12] On the other hand, the unbound or distant planetary mass population detected by gravitational micro-lensing has been reported.[13] Of the various methods targeting exoplanets, the observation of exoplanets spatially resolved from their parent star and monitoring transiting exoplanets are potentially suitable for spectroscopic characterization of their atmospheric features, including biomarkers. Both these methods are "high dynamic-range" observations.

The Space Infrared telescope for Cosmology and Astrophysics (SPICA)[14] has the potential to provide a unique opportunity for high dynamic-range observations of exoplanets because of its large telescope aperture, high stability, and the capability for making wideband infrared observations from deep space. The SPICA Coronagraph Instrument (SCI)[15] is a proposed high-dynamic range instrument especially for targeting exoplanets. In this paper, first, reviews of the SPICA mission and the SCI are presented, and then technical progress and ideas in the conceptual studies and the technology are shown.

---

*enya@ir.isas.jaxa.jp

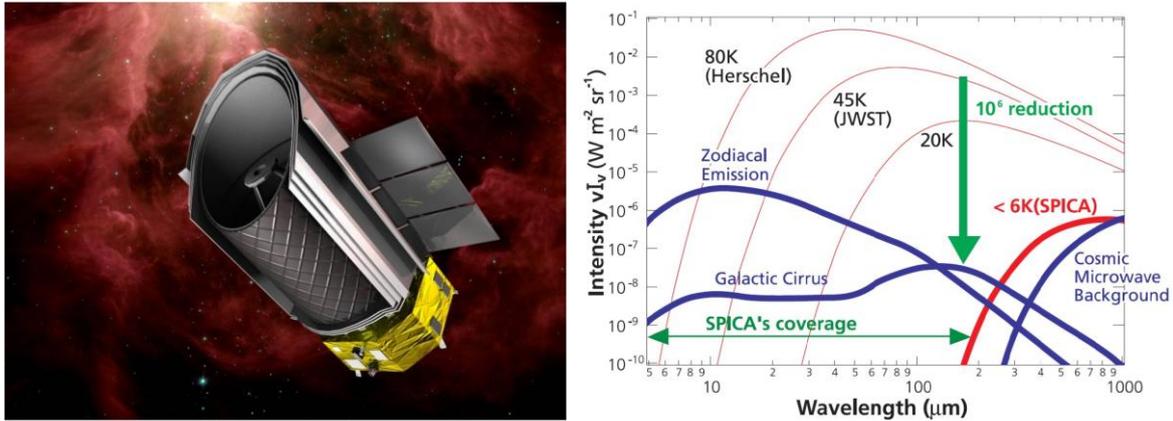

Figure 1. (Left) artistic view of SPICA in orbit. (Right) comparison of the natural background (zodiacal emission, Galactic cirrus, and cosmic microwave background) with those of the thermal radiation from telescopes as a function of temperature.

## 2. THE SPICA MISSION

The Space Infrared telescope for Cosmology and Astrophysics (SPICA) is a mission optimized for mid-infrared (MIR) and far-infrared astronomy with a 3 m class telescope (3.0 m effective pupil diameter, 3.2 m physical diameter in the current design).[14] Artistic view of SPICA in orbit is shown in Figure 1. With its unique capabilities, SPICA is expected to provide an essential opportunity to study some important issues in astrophysics, e.g., the birth and evolution of galaxies, the formation and evolution of stellar and planetary systems, the chemical evolution of the universe, and so on. SPICA is planned to be launched into the Sun-Earth L2 Halo orbit in the fiscal year (FY) 2018. The whole telescope, together with the focal plane instruments, will be launched at ambient temperature and then cooled to less than 6 K in orbit in order to achieve ultra-high sensitivity in the infrared[16] as shown in Figure 1. On-axis Ritchey-Chrétien optics are used for the telescope[17,18] which is designed to be diffraction limited at 5 μm wavelength and have a wave front error (WFE) < 350 nm rms. Sintered SiC and hybrid C/SiC are Candidate material for the mirrors and the structure of the telescope. The guaranteed lifetime is 3 years, with the goal of extending operations to 5 years. The proposed focal plane instruments for SPICA and outlines of their functions in the current design are as below (see also Figure 2).

**MIR Camera and Spectrometer (MCS):** This is a general purpose instrument for MIR imaging and spectroscopy, covering the wavelength region of 5-38 μm with Si:As and Si:Sb detectors.[19,20] The MCS consists of several channels: a wide field camera (WFC) with a field of view of 5'×5', a High-Resolution Spectrometer (HRS) with a resolution of R~30000, a Medium-Resolution Spectrometer (MRS), and a Low-Resolution Spectrometer (LRS).

**SPICA FAR-infrared Instrument (SAFARI):** This instrument covers the ~34-210 μm waveband with a spectral resolution of R ~10 to 10 000, and a field of view of 2' × 2'. A far-IR imaging FTS-spectrometer of SAFAR is especially valuable with SPICA. SAFARI is proposed by a consortium of European and Canadian institutions led by SRON.[21,22,23]

**Far-infrared and sub-millimeter spectrometer:** NASA issued an AO for the study of sensitive far-infrared and sub-millimeter spectrometers for SPICA in 2009, and three teams were selected for this study.[24,25,26]

**Focal Plane Camera (FPC):** The FPCs (the guider camera, FPC-G, and the science camera, FPC-S) will be procured by KASI as a PI institute. Both the FPC-G and the FPC-S adopt InSb detectors and cover a field of view of 5'×5'. The FPC-G and the FPC-S are used in the I-band (0.8μm band) and the wavelength region of 0.7-5μm, respectively.[27]

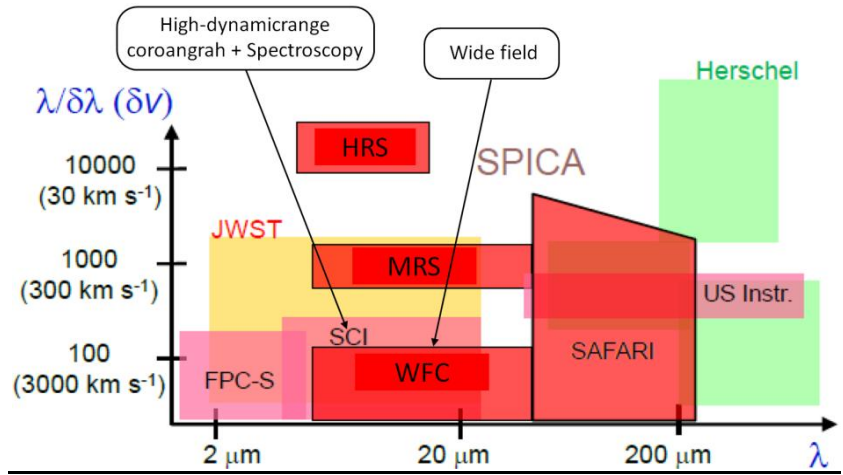

Figure 2. Overview of scientific instruments onboard SPICA.

**SPICA Coronagraph Instrument (SCI):** The coronagraph is used in the MIR for high contrast observations by suppressing the side lobes of the Point Spread Function (PSF) of a bright source (i.e. a star) enabling observation of companion objects (e.g., planets and proto-planetary disks).[15,28] The coronagraph works in both imaging and spectroscopy modes, whilst a non-coronagraphic mode is also available. A more detailed description is given in Section 3.

The SPICA spacecraft consists of a cryogenic Payload Module (PLM) and a Bus Module (BM). The PLM includes the Cryogenic Assembly and the Scientific Instruments Assembly. The Cryogenic Assembly includes mechanical coolers and passive thermal shields, which are required to cool down and maintain the Scientific Instruments Assembly to less than 6 K.[16] The latter includes the SPICA Telescope Assembly (STA) and the Instrument Optical Bench (IOB). Additional coolers enable the temperature of the focal plane detector to be cooled to ~ 100 mK. The overall PLM architecture is dominated by the need to have a large Sun shield, three coaxial thermal shields, a telescope shell and an additional telescope baffle. The PLM and BM are connected by a low thermal conductivity truss structure, which also supports the STA. The BM has a conventional design and includes most of the spacecraft subsystems (power, propulsion, attitude control, data handling, thermal control and telecommunications). The total mass of the spacecraft is about 3.7 tons, in line with the capabilities of the baseline launcher vehicle, H IIA-204 of JAXA. The Sun - Earth L2 point is the optimum environment to obtain wide sky visibility and a stable thermal environment to enable cooling of the telescope. The absence of cryogens onboard allows the lifetime to be extended beyond its nominal value.

The thermal environment for the telescope and the instruments is maintained by a combination of passive and active cooling. The former is done by dedicated Sun and thermal shields combined with radiators, and the latter by a number of mechanical coolers, ensuring base temperatures of 4.5 K and 1.7 K. SPICA has adopted a new concept in cryogenic systems that do not use cryogens, as shown in Figure 3. The elements, including the STA, the IOB and some focal plane instruments, are maintained at 4.5K by the combined action of mechanical cooling and efficient radiative cooling in the stable thermal environment at the Sun-Earth L2 point. For the thermal design, the cooling capacity of a 4 K-class mechanical cooler at its end of life was considered, and it was assumed that the Joule heating of the focal plane instruments would be approximately 15 mW. The baseline design of the thermal insulation and radiative cooling system was determined by thermal and structural analyses, so that the total parasitic heat flow from the higher temperature elements to the 4.5 K stage would remain below 25 mW. The Mechanical Cooling System of SPICA is based on a set of advanced Stirling and Joule-Thompson coolers. The 4K Mechanical Cooler for the 4.5 K stage is a 4K-class Joule-Thomson cooler connected to a pre-cooler comprising a 20 K-class two-stage Stirling cooler. Far-infrared instruments such as SAFARI require further cooling to 1.7 K. Such temperatures are realized with a 1K Mechanical Cooler consisting of a dedicated 1K-class Joule-Thomson cooler with $^3$He working gas and an additional two-stage Stirling pre-cooler. Previous and ongoing Japanese missions, e.g., AKARI[29], SMILES[30], and Astro-H[31], were equipped with similar mechanical cryo-cooers and such missions have left a large heritage for SPICA.

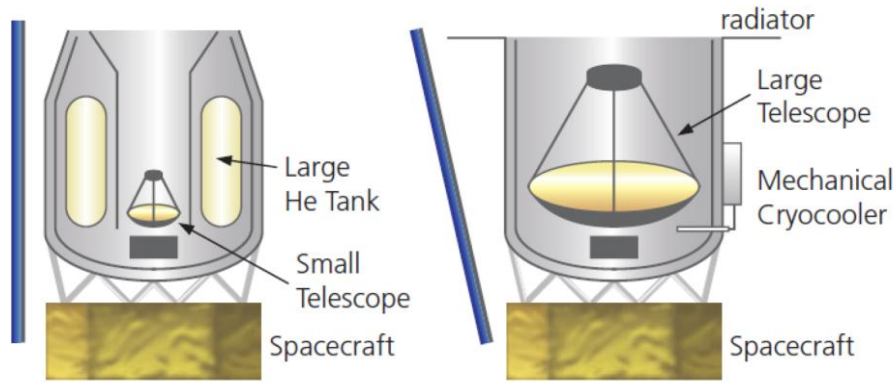

Figure 3. Comparison of conventional (left) and SPICA's cryogenic design (right).

JAXA will develop MIR instruments together with universities and research institutes in Japan. The SPICA mission proposal was submitted to JAXA in 2007, following extensive discussions in the community for more than 10 years in Japan. After several reviews, the SPICA pre-project team was formed at JAXA in July 2008, and the three-year phase-A activity started. The SPICA phase-A study consists of two parts, the concept design and the project formalization. The first part of the pre-project, and the system requirements review were finished. After this review, the latter half of the phase-A activity was started. At the end of this phase, SPICA is required to go through another technical review, the system definition review. For the proposed focal plane instruments, the review in Japan by the SPICA task force and the review committee has been completed, and the international review is ongoing. Following these reviews, the Project Phase-up Review, which is a management review by JAXA, is planned, toward the launch of SPICA in FY 2018.

SPICA is an international mission in which ESA is an essential partner.[32,33] The European SPICA Consortium submitted a proposal to enable European participation in SPICA to ESA in June 2007 under the framework of the ESA Cosmic Vision 2015-2025. The proposal called on ESA to assume a partner agency role in SPICA by making contributions to the SPICA Telescope Assembly, the European SPICA Ground Segment, SPICA SAFARI I/F Management, and SPICA Mission user support. The proposal also assumed that SAFARI was to be procured by the European Consortium. The proposal was selected by ESA in October, 2007 as one of the candidates for future missions. Then an assessment activity on SPICA led by ESA was conducted from November 2007 to August 2009. Following the study, another proposal for the next phase was submitted to ESA in September 2009. The scientific goals of SPICA were highly regarded by the Space Science Advisory Committee and the Science Program Committee. Transition to an implementation phase will require approval by the SPC.

The possibility that the US procures one of focal plane instruments has been discussed extensively. NASA issued an AO for the study of sensitive far-infrared and sub-millimeter spectrometers for SPICA in 2009, and three teams were selected for this study. They are now making an extensive assessment study. Korean participation in SPICA has also been discussed extensively. The current plan is that the FPCs (both FPC-G and FPC-S) will be procured through KASI as a PI institute.

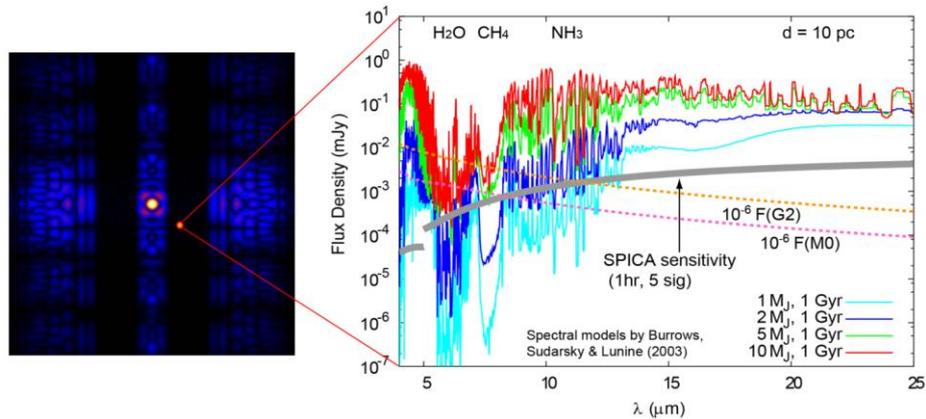

Figure 4. Coronagraphic image (left) and spectrum (right) expected from observations with the SCI. The planet spectrums are from Burrows, Sudarsky, & Lunine (2003).

## 3. SCI

The SCI is an instrument designed especially for the extensive study of exoplanets.[15] SPICA has advantages as the platform for the SCI: it is perfectly free from band pass limitations and wavefront turbulence caused by the air. The cooled telescope provides high sensitivity in the infrared region. High stability is expected since the cryogenic telescope is to be launched into deep space, into the Sun–Earth L2 Halo orbit. The instrument is designed to carry out two "critical tasks". One of the critical tasks is the coronagraphic detection and characterization of Jovian exoplanets (Figure 4). The other is the monitoring of transiting planets, which is based on the non-coronagraphic mode of the SCI with simultaneous wide wavelength coverage by two channels, the short-and long-wavelength channels, as described below. The "given" performance of the designed instrument will allow other important scientific studies to be carried out with the SCI.

Figure 5 shows the optics of the SCI. Total mass of the SCI is 20 kg. The SCI is cooled in orbit before operation, together with the other focal plane instruments and the telescope. The SCI uses the center of the field of view in the focal plane of the telescope to obtain the best wavefront quality. All the mirrors for collimation and focusing in the SCI are off-axis parabolas. Aberration is designed to be minimized at the center of the field of view in order to obtain the PSF of the parent star.

The pre-coronagraphic optics and the coronagraphic optics are common for both wavelength channels, whilst the post coronagraphic optics is split into two channels. The pre-coronagraphic optics consists of reflective optical devices in order to avoid ghosts and/or wavelength dependence. In contrast to the other focal plane instruments of SPICA, the SCI has no flat pick-off mirror, and the beam from the secondary mirror of the telescope arrives directly on the collimating mirror. The first mirror of the SCI is placed after the focal point of the telescope. This is convenient for optical tests with the SCI and the telescope. When a deformable mirror (DM) is adopted to correct the wavefront errors of the telescope[34,35], it will be installed at the pupil plane re-made in the SCI.

The baseline design of the SCI adopts a binary-shaped pupil mask coronagraph due to peculiar requirements to SPICA[15,36]: first, the SPICA coronagraph has to work in the wide MIR wavelength region in a cryogenic environment, so achromatism is an important property, and a coronagraph without transmissive devices is preferable. Second, the coronagraph should not be subject to telescope pointing errors caused by vibration of the cryogenic cooling system and the attitude control system. Furthermore, simplicity, compactness, and being lightweight are also considered. Binary-shaped pupil-mask coronagraphs are in principle achromatic (except for the PSF scaling with wavelength) and robust against telescope pointing.[37-49] In this coronagraph, a single binary pupil mask modifies the PSF and provides the required contrast. Some binary shaped pupil masks were designed for the SPICA pupil with the obscuration as illustrated in Figure 5. These designs consist of multi-barcode masks "skipping over" the obscuration and have coronagraphic power in one dimension only. As a result, a large opening angle is realized in keeping with the specification for the Inner Working Angle (IWA). In the masks shown in Figure 5, mask-1 is the baseline design. Additionally, mask-2 provides a small IWA to explore the field closer to the parent star. On the other hand, the contrast of mask-2 is not as high as mask-1 because there is a trade-off between the IWA and the contrast. These two masks are complementary, and can be changed by a mechanical mask changer. The mechanical changer is also essential to provide the non-coronagraphic mode to the SCI. It should be noted that the principle of the barcode mask was presented by Kasdin et al. (2005)[43], and the LOQO optimizer presented by Vanderbei (1990)[50] was used for optimization in these designs. More details about these designs are given in references.[15, 36] Further new improvements are described in Section 4.1. A focal plane mask is practically used to obscure the bright core in the coronagraphic PSF and to prevent scattered light from the PSF core polluting the dark region of the coronagraphic PSF in the post-coronagraphic optics. Interchangeable focal plane masks will be used to realize different observing modes, including a slit that can be used both with and without the pupil mask to provide spectroscopic capability in coronagraphic and non-coronagraphic observing modes.

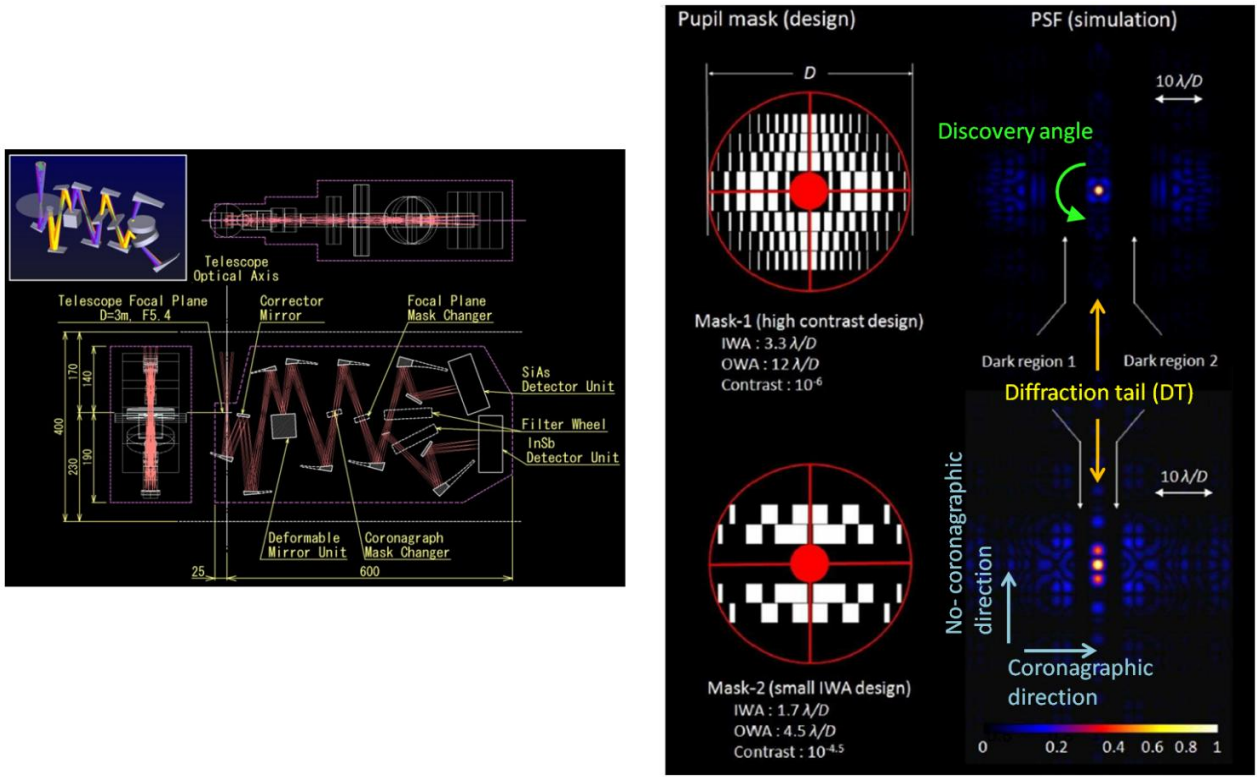

Figure 5. (Left) optical layout in the SCI. (Right) pupil masks and PSFs produced by the masks.

The post-coronagraphic optics uses a beam splitter to split the optical path into two channels, the short wavelength channel with an InSb detector and the long wavelength channel with a Si:As detector. Each channel has filter wheels, and simultaneous observation of the same target with the two channels is possible. Each filter wheel contains a transmissive disperser (e.g., grisms, prisms) for spectroscopy.

In order to reduce the technical risk in the development, the design of the instrument without the use of a DM is regarded as the baseline design. In parallel with the baseline design, an advanced design with a DM is being studied. As a result of this simplification, the contrast at the PSF is limited to $10^{-4}$. In this case, the advantage of high contrast over the JWST is basically lost. However, the ability of coronagraphic spectroscopy is kept as a unique capability of the SCI. Young outer planets observed by direct imaging have already been reported.[8,9,10] These planets are sufficiently bright in the infrared region for spectroscopic observations using the simplified design of the SCI. The spectral data of such targets in the wide infrared wavelength range is essentially unique and important.

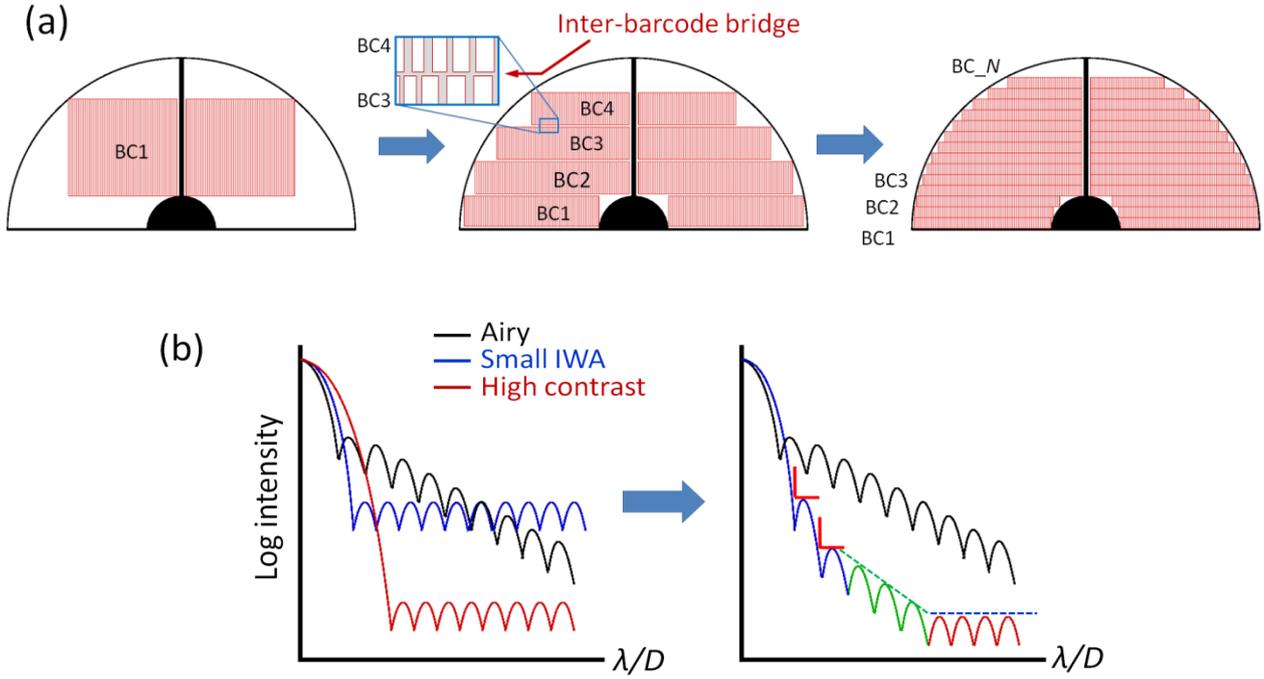

Figure 6. (Top) Schematic view for the integral barcode masks for a 1-D coronagraph. (Bottom) Concept of generalized constraints on the darkness of the PSF halo.

## 4. STUDIES FOR THE DESIGN AND DEVELOPMENT

### 4.1 Mask design

This section presents the results of a conceptual study to evolve the mask design. As shown in Enya and Abe (2010)[36], it is possible to "skip over" the pupil obscuration with a binary shaped pupil mask, and it is also possible to optimize each barcode of a mask independently, as illustrated in Figure 6 (a). Based on this fact, the integral one dimensional (1-D) binary pupil-mask coronagraph was derived. Thanks to using the dead area of the pupil, the barcode can be expanded and then the absolute throughput of the barcodes is improved. Furthermore, the design requirements, e.g., those for the IWA, can be relatively relaxed for each expanded barcode. So the relative throughput of each barcodes is also increased. As a result of these effects, it is expected that the efficiency of the mask will be significantly improved. In the case of a 1-D binary pupil-mask coronagraph, an "inter-barcode bridge" is applicable between two neighboring barcodes. A mask consists of rectangular holes which have no corner with very acute angle. These facts are advantageous for the manufacturability of free standing masks.

Figure 6 (b) presents the concept of the generalized darkness constraint. The left panel corresponds to a previous design, in which the IWA, the Outer Working Angle (OWA), and the contrast are given as the constraints and optimization was done to maximize the throughput of the mask. In this case, the IWA and the contrast are in a strong trade-off relationship, i.e., requiring a small IWA limits the contrast to relatively low values, whilst a high contrast requirement means the IWA has to be relatively large. A lower contrast than the Airy profile was even possible. In the right panel is an illustration of the generalized constraint consisting of two sets of IWA and contrast (red marks), slope constraints (dashed green line), and flat constraints (dashed blue line). A flat constraint area can be useful to avoid pushing the contrast requirement higher than other limits, e.g., telescope sensitivity. Tests on these systems have brought an understanding that an especially strong trade-off relationship exists between the IWA and the contrast close to the PSF core, rather than in the outer region. Thus, it turns out that there is more flexibility in improving the efficiency of the mask design.

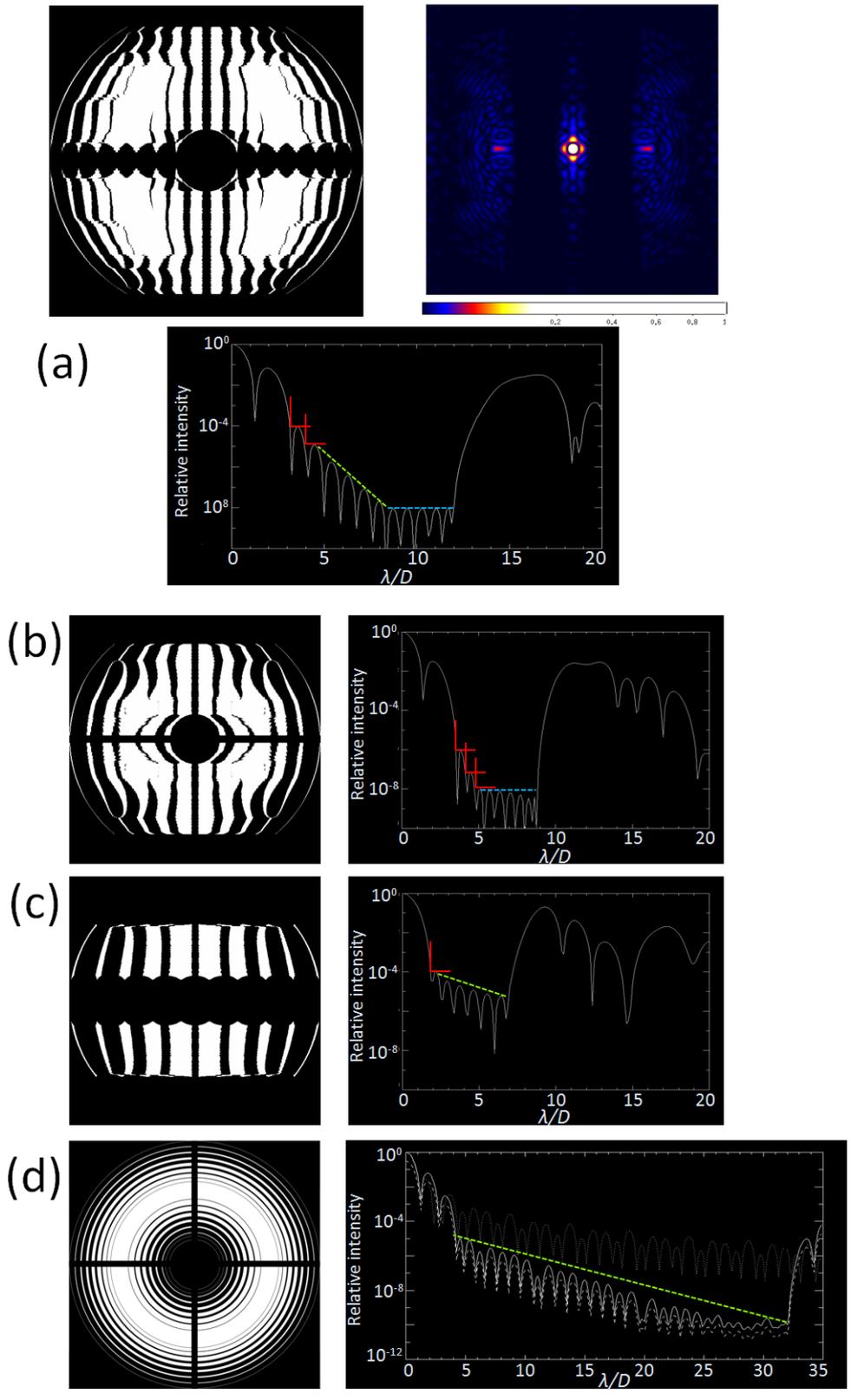

Figure 7. Examples of mask designs.

Figure 7 (a) presents an example design of the integral 1-D binary pupil-mask coronagraph with general darkness constraints. The number of grids defining each barcode is 512 over the pupil. For simplicity, the inter-barcode bridge is not included. It should be noted that the LOQO optimizer presented by Vanderbei (1990)[50] was used for optimization of each barcode for this design, as well as all designs shown in this paper. The top-left, top-right, and bottom panels are the mask design, the calculated PSF with the mask design, and the profile of the PSF in the coronagraphic direction, indicated in Figure 5. Its contrasts and IWAs are $10^{-4}$ at 3.2 $\lambda /D$, and $10^{-5}$ at 3.8 $\lambda /D$, respectively. These values of contrast are in a suitable range for use in the case of SPICA without adopting a DM. The flat constraints area with a contrast of $10^{-8}$ spreads between 8.5 and 12 $\lambda /D$. The throughput of this mask is 50 %. Figure 7 (b) shows a high-contrast solution. Its contrasts and IWAs are $10^{-6}$ at 3.6 $\lambda /D$, $10^{-6}$ at 4.1 $\lambda /D$, and $10^{-8}$ at 4.7 $\lambda /D$, respectively. These high contrasts are potentially suitable for observation with wave front control with a DM. The throughput of this mask is 41 %. A solution intended for a small IWA is shown in Figure 7 (c). The contrast for this solution is $10^{-4}$ at an IWA= 1.8 $\lambda /D$, and $10^{-5}$ at an IWA = 7.0 $\lambda /D$, respectively. The throughput of this mask is 23 %. For both designs in Figure 7 (b) and (c), the number of grids defining each barcodes is 512 over the pupil, and the inter-barcode bridge is not given.

Figure 7 (d) presents a trial design for a large OWA for a wide field survey. This design is based on a concentric ring mask[39] with obscuration by the secondary mirror and the support structure of the telescope[44], but with generalized darkness constraints. The optimization was carried out with the central obscuration. Then obscuration by the support structures was added. Its IWA is 4.0 $\lambda /D$, and the contrast there is $10^{-5}$. A simple slope constraint is applied, which reaches a contrast of ~$10^{-10}$ at an OWA=32 $\lambda /D$. This solution indeed proves that the requirement of high contrast in the outer area is not demanding. Because of the support structure, 4 diffraction tails (90 degree symmetry) appear at the PSF. These diffraction tails prevent efficient observation very close to the PSF core. On the other hand, the influence of the diffraction tails is less serious in the outer region of the PSF than in the inner region, or practically almost negligible. Therefore, this mask is complementary to the integral 1-D binary pupil-mask. The masked area corresponding to the support structure works as an "inter-ring bridge" which helps the manufacturability of a free standing mask of this design. The throughput of this mask is 53 %.

The designs presented in this section are the result of a conceptual study and are just examples. Other various combinations of constraints are possible, and need to be researched, with consideration given to scientific simulations and the manufacturability of the mask.

### 4.2 Spectral disperser for monitoring transiting exoplanets

As described in Section 1, the monitoring of transiting exoplanets is a critical issue for a space infrared telescope. For this observation, it is important to design the instrument to be complementary to the James Webb Space Telescope (JWST).[51,52,53] The non coronagraphic mode of the SCI provides us with the capability of simultaneously monitoring in the short and long wavelength channels. Such a function is not available to the other SPICA instruments or the JWST. This section presents a conceptual study of the optics needed for spectroscopic measurements to realize wide wavelength coverage in each channel.

Figure 8 (a) presents an example of a design which covers the full wavelength region of the long channel of the SCI. The left and the right panels show the optical layout and the spectral format, respectively. Three prisms made of Kbr, CsI, and KBr are used to achieve broad wavelength coverage and a straight optical path. In the spectral format, the black spots and red crosses show the spot diagram and the Airy disk size, respectively (diffusion of the PSF by the wavefront error of the telescope is not reflected). A three prism system is used to obtain higher spectral resolution, whereas a two prism system has the advantages of high throughput, less optical ghosts, and simplicity. The preferable spectral dispersers for the SCI are to be simple units which work only in one position in the imaging optics. The solution of Figure 8 (a) is just one such unit designed for installation into one slot of the filter wheel at the pupil plane, though the prism unit is somewhat thick. It is in principle possible to obtain higher spectral resolution by the cross disperser unit using prisms with gratings made on the surface of the prisms, as Figure 8 (b). The material used for this design is Kbr, CsI, and KBr, too. In this case the primary role of the prisms is only to separate the order of the spectrum, and the spectral resolution is mainly provided by the grating. Figure 8 (c) shows a

## Optical Layout

### (a) Long Channel : Direct Vision Prism

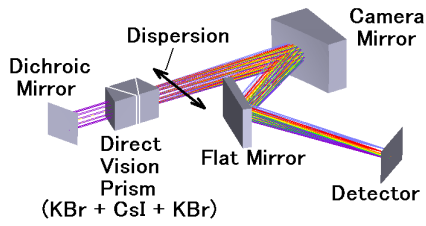

### (b) Long Channel : Cross Dispersion by Prism + Grism

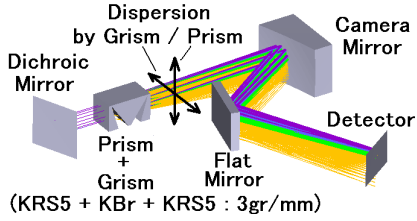

### (c) Long Channel : Direct Vision Prism
for $\lambda$ = 5-13 $\mu$m

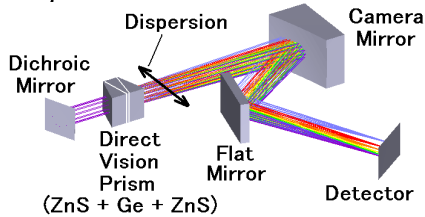

### (d) Short Channel : Direct Vision Prism

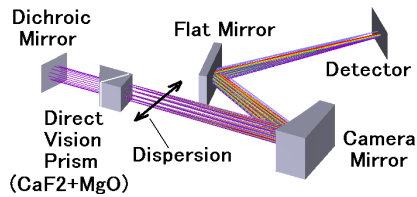

### (e) Short Channel : Cross Dispersion by Prism + Grism

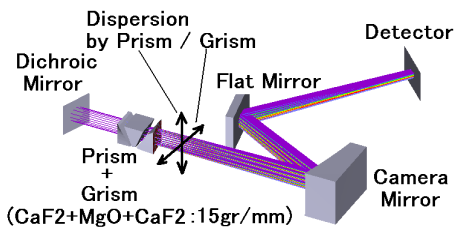

## Spectral Format

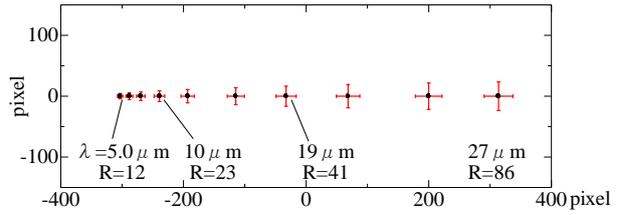

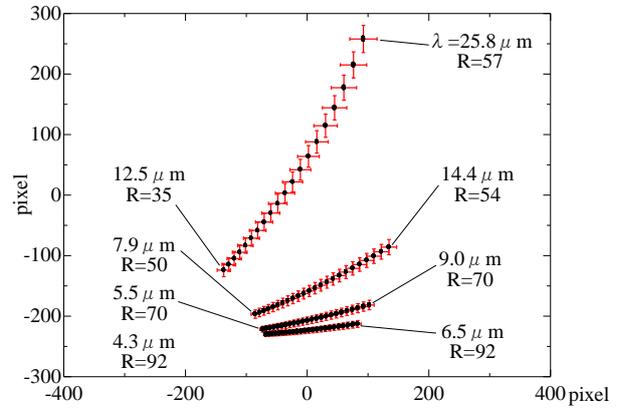

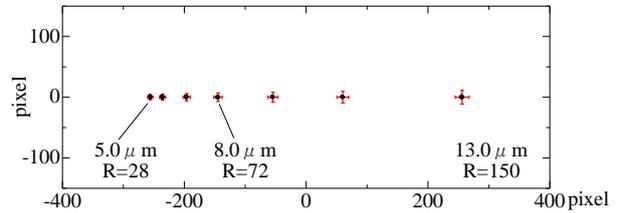

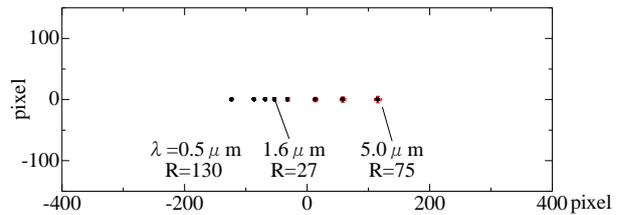

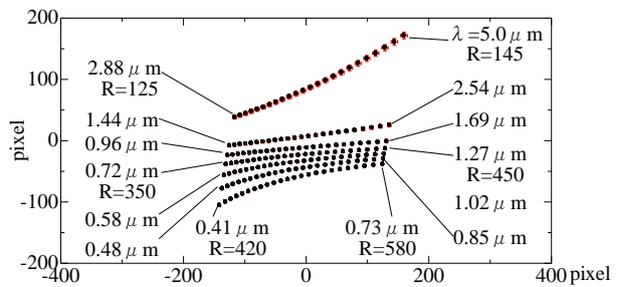

Figure 8. Trial design of spectral disperser for wide wavelength coverage.

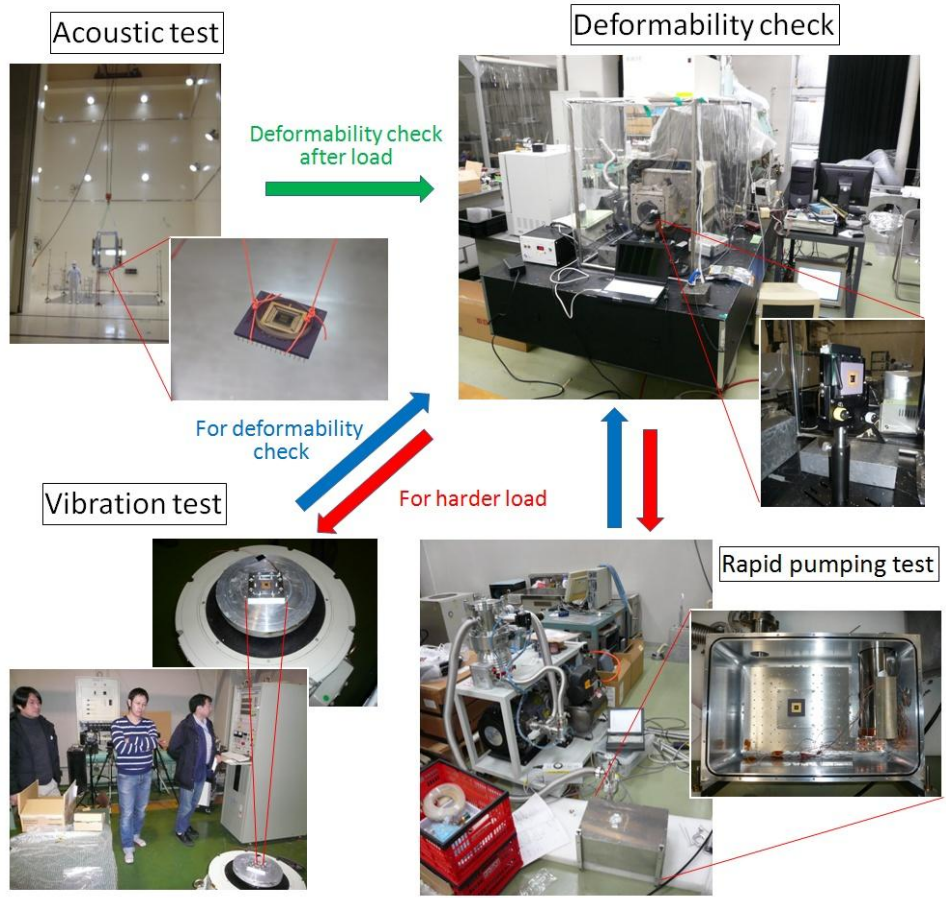

Figure 9. Toughness tests for MEMS DMs.

simple solution consisting of three prisms without gratings adapted for the wavelength range of 5-13μm. As the result of a compromise on the wavelength coverage, higher spectral resolution than the design of Figure 8 (a) was obtained. The material used for this solution is ZnSe and Ge as well as the prisms used in the JWST, whilst the design consisting of two prisms was adopted for the JWST. A similar design is possible for the short channel of the SCI. Figure 8 (d) and (e) are example solutions with two prisms, and cross dispersion by prisms and gratings, respectively. $CaF_2$ and MgO are used for the materials of the prisms.

The designs presented above are the result of a conceptual study and are just examples. Other designs are also possible; for instance, it is possible to give curvature (e.g., cylindrical) to the surfaces of the prisms, or to add an optical device with curvature (e.g., a lens). Such a device works as a diffuser to avoid photon saturation on the detector in the observation of bright targets. In any case, optimization is needed for the material selection, the number of prisms, the geometry of the devices, and the grating design. Manufacturability is also an important issue, especially for the feasibility of making gratings on each material for the case of cross dispersion. Finally a trade-off study should be done simulating the scientific observations.

### 4.3 Toughness tests of the DM

If a DM is applied to correct the wavefront error, the contrast of the coronagraph is vastly improved and so the development of a DM is quite valuable. A DM made by Micro Electro Mechanical Systems (MEMS) comprises a micro-cavity structure with a membrane surface and small electrical pads, and is actuated by Coulomb forces.

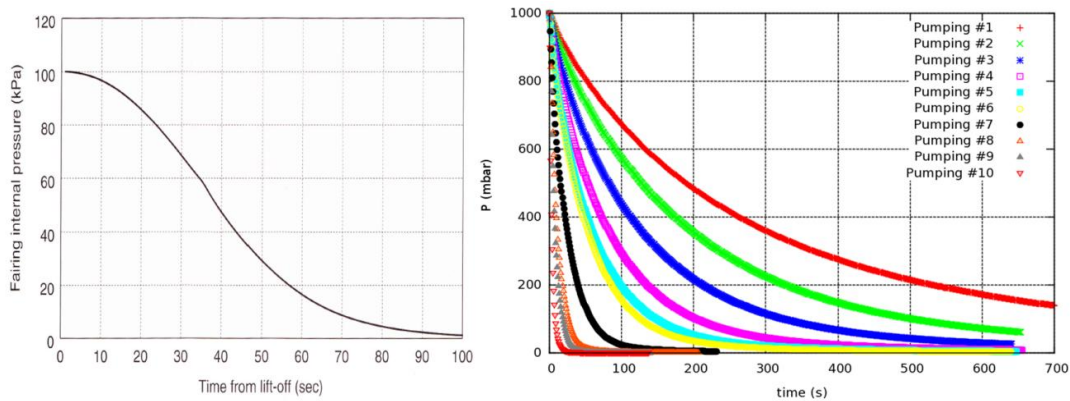

Figure 10. Pressure profiles at the fairing of the H II-A rocket (left), and experimental data obtained in the DM tests (right).

MEMS DMs have been studied for applications in a space cryogenic telescope[34,35] because they are compact, light weight, have a large format, and in principle the actuation property is less dependent on temperature (c.f., DM by piezoelectric actuation). On the other hand, it is not clear if the microstructure of a MEMS DM is tough enough to survive the various loads; acoustic, vibratory, and the rapid vacuum pumping at the launch of the rocket. Thus, a toughness test for MEMS DMs was recently started. This section summarizes the first results of the test.

First, the opportunity for an acoustic test arose, as a result of an acoustic test to be done on the solar activated paddles of JAXA's small satellite, SPRINT-A, at Tsukuba Space Center of JAXA on Feb. 3rd, 2011. To exploit this opportunity, a commercially available MEMS DM with 32 channels made by Boston Micromachines Corp. (BMC) was used, rather than developing a special device by order. The DM was placed in the acoustic test room (the top-left of Figure 9). The acoustic load level is described in Table 1. The load level was defined for the launcher of SPRINT-A, an Epsilon rocket of JAXA. As the result, the applied load level is slightly different from the level of an H II-A rocket. The load was maintained for 60 second. After the load, a detailed visual inspection of the DM and a deformability test using an interferometer, ZYGO GPI-XP were carried out in the Institute of Space and Astronautical Science (ISAS), JAXA. As the result, it was confirmed that there was neither significant destruction nor any changes in the deformability properties of the surface.

Next, vibration tests were executed on Feb. 14th, 2011 with the DM used for the acoustic test. The DM was installed on a small vibration stage in ISAS (bottom-left of Figure 9). The vibration produced by this instrument was only in the vertical direction. The zero-db level of the vibration load is described in Table 2. Firstly, a load of -12 db level was applied for 60 second, and then a detailed visual observation and the deformability measured by the interferometer were applied in ISAS. The same processes were carried out for -6db, -3db, 0db, and +3db load levels. The strongest load corresponds to a significantly higher level than that expected at the rocket fairing at the actual launch. As a result of the tests, it was confirmed that there was neither significant destruction nor any change in the deformability properties of the surface.

A rapid vacuum pumping test was also pursued (bottom-right of Figure 9) in ISAS. For this test, a DM with 32 channels but without any cover window was made by BMC through a special order. With the exception of the window, this DM was comparable to the one used for the acoustic and vibration tests. The pumping speed was reduced by using a resistance to the airflow installed in the pumping tube. Various resistances were used to realize ten different pumping pressure profiles (the left panel of Figure 10). The right panel of Figure 10 shows the profile of the pressure decreasing at the rocket fairing of H II-A.[54] For the first experiment, the most modest pumping with the largest resistance, represented by profile #1 in Figure 10, was applied to the DM set in a vacuum chamber. Next, the DM was taken from the chamber, and a detailed visual inspection was carried out and the deformability measured by an interferometer. The same processes were carried out with more rapid pumping step by step, i.e., profiles of #1, #2, … and finally #10 (Figure 10). In all cases, the chamber leakage was less than the pressure changes in profile #1. As a result, it was confirmed that there was neither significant destruction nor changes in the deformability properties of the surface. It should be noted that the pumping of #10 is more rapid than the rate at which the pressure decreases in the actual rocket fairing over any pressure range.

Table 1. Load levels used for the acoustic test.

| 1/1oct center frequency | Acoustic pressure (dB) | tolerance |
|---|---|---|
| 31.5 | 129.0 | +3/-10 dB |
| 63 | 135.0 | +- 3dB |
| 125 | 141.6 | +- 3dB |
| 250 | 138.0 | +- 3dB |
| 500 | 134.2 | +- 3dB |
| 1000 | 129.0 | +- 3dB |
| 2000 | 126.0 | +- 3dB |
| 4000 | 121.0 | +3- 10dB |
| 8000 | 117.0 | +- 6dB |
| Over all | 144.6 | +- 2dB |

Table 2. The 0-db load level used for the vibration test.

| Frequency (Hz) | PSD (G$^2$/Hz) |
|---|---|
| 20 | 4.3 |
| 80 | 67.3 |
| 270 | 67.3 |
| 413 | 28.9 |
| 800 | 28.9 |
| 2000 | 2.5 |
| Over all | 21.1 Grms |

Instead of the use of small DMs with 32 channels, it is important to manufacture devices close to the flight model. Other essential issues are to estimate and use the acoustic, vibrational, and rapidly decreasing pressure load levels derived from considering the device set in the SCI installed in the rocket fairing together with whole the satellite, and to carry out load tests with more realistic load levels. For this purpose, it is needed to fix the designs of the instruments and of the spacecraft, and to complete a detailed analysis and simulation of the structures. Although the DM and the tests described above were not the final ones, but just initial ones, it was encouraging that the DM survived all of the acoustic, vibratory, and vacuum pumping tests, and also the earthquake on Mar. 11th, 2011.

As a potentially possible backup, it is interesting to consider the concept of a passive corrector mirror using no actuator. The SPICA telescope will be tested in a cryogenic chamber with an interferometer. It is expected that the wavefront error of the telescope after launch will be estimated based on data from cryogenic measurements of the surface figure and numerical simulations for gravity release. The passive corrector mirror should be polished to have a surface figure which cancels the predicted wavefront error of the telescope. For this method, consistency of the timeline between manufacture of the corrector mirror and the whole project is a critical issue, as well as the accuracies and limits in the predicted wavefront error of the telescope, the polishing, the measurement, and the alignment and so on, which can depend on the size of the mirror.

## ACKNOWLEDGEMENT

We give our heartfelt thanks to all the pioneers in this field, all our colleagues, and all those related to this work, especially to R. Vanderbei, J. Kasdin, P. Bierden, S. Cornelissen, M. Feinberg, and C. Lam. The work presented in this paper is supported by JAXA and Grants-in-Aid for Scientific Research/The Ministry of Education, Culture, Sports,Science and Technology of Japan.